# Difference Autocorrelation: A Novel Approach to Estimate Shear Wave Speed in the Presence of Compression Waves

Hamidreza Asemani*, Jannick P. Rolland, and Kevin J. Parker

*Abstract:* In share wave elastography (SWE), the aim is to measure the velocity of shear waves, however unwanted compression waves and bulk tissue motion pose challenges in evaluating tissue stiffness. Conventional approaches often struggle to discriminate between shear and compression waves, leading to inaccurate shear wave speed (SWS) estimation. In this study, we propose a novel approach known as the difference autocorrelation estimator to accurately estimate reverberant SWS in the presence of compression waves and noise. *Methods:* The difference autocorrelation estimator, unlike conventional techniques, computes the subtraction of velocity between neighboring particles, effectively minimizing the impact of long wavelength compression waves and other wide-area movements such as those caused by respiration. We evaluated the effectiveness of the integrated difference autocorrelation (IDA) by: (1) using k-Wave simulations of a branching cylinder in a soft background, (2) using ultrasound elastography on a breast phantom, (3) using ultrasound elastography in the human liver-kidney region, and (4) using magnetic resonance elastography (MRE) on a brain phantom. *Results:* By applying IDA on the unfiltered contaminated wave fields of simulation and elastography experiments, the estimated SWSs are in good agreement with the ground truth values (i.e., less than 2% error for the simulation, 9% error for ultrasound elastography of breast phantom and 19% error for MRE). *Conclusion:* Our results demonstrate that IDA accurately estimates SWS, revealing the existence of a lesion, even in the presence of strong compression waves. *Significance:* IDA exhibits consistency in SWS estimation across different modalities and excitation scenarios, highlighting its robustness and potential clinical utility.

*Keywords:* Difference autocorrelation, magnetic resonance elastography, reverberant shear wave, shear wave elastography, ultrasound elastography.

## I. INTRODUCTION

Shear wave elastography (SWE) is an expanding imaging technique with applications in different modalities such as ultrasound [1,2], magnetic resonance imaging (MRI) [3,4], and optical coherence tomography (OCT) [5,6]. The capability of this method to estimate tissue stiffness and highlight different lesions positions it as a robust clinical tool for diagnosing a broad spectrum of diseases [7,8]. The fundamental concept of SWE remains consistent across different imaging modalities. Typically, shear waves are induced in the tissue through external excitation captured by the imaging modality and processed to compute the shear wave speed (SWS) map or tissue stiffness map [9-12].

The resolution and accuracy of SWE are dependent on the imaging modality, the characteristics of the generated shear wave field, and the post-processing techniques applied [10,13].

In SWE, the focus lies on measuring the speed of shear waves rather than compression waves, as unwanted compression waves and translational motions can introduce challenges in accurately assessing tissue stiffness. As illustrated in the classic text by Graff [14], a vibrating source on the surface of a body will impart shear waves, which are useful for elastography, but also significant compression waves and surface waves, which can confound the estimate of shear wave speed. Compression waves and any other bulk tissue motion have long wavelengths and long correlation lengths compared to shear waves, so they present an unwanted term if the analysis is oriented toward shear waves. Generally, these have been minimized by post-processing steps, including the calculation of the vector curl from 3D data or more simply from highpass filtering of 2D displacement data [15].

*H. Asemani is with the Institute of Optics and the Department of Electrical and Computer Engineering, University of Rochester, Rochester, NY, USA (correspondence e-mail: hasemani@ur.rochester.edu). J. P. Rolland is with the Institute of Optics and the Department of Biomedical Engineering, University of Rochester, Rochester, NY, USA. K. J. Parker is with the Department of Electrical and Computer Engineering and the Department of Biomedical Engineering, University of Rochester, Rochester, NY, USA.

An example of filtering is given in [16], where Ormachea and Parker employed a 2D bandpass spatial filter to eliminate extremely low spatial frequency compressional waves and reduce high frequency noise in all directions. They used the cutoff spatial frequencies ($f$) related to the wavenumber $k$ of the filter, which were set at $k_l = 2\pi f/C_h$ and $k_h = 2\pi f/C_l$, where $C_l$ and $C_h$ represented the chosen low and high SWSs, respectively.

In this study, we introduce the difference autocorrelation estimator in order to calculate SWS in the presence of compression waves. By subtracting the velocity between two neighboring particles, we effectively minimize the impact of compression waves. This technique is capable of estimating SWS in fully reverberant shear wave fields, as well as imperfect or more directionally oriented shear wave fields. The application of the proposed approach is studied using (1) k-Wave elastography simulation of a stiff branching cylinder in a soft background, (2) ultrasound elastography of a breast phantom with a lesion (3) ultrasound elastography of the human liver kidney region, and (4) magnetic resonance elastography (MRE) of a brain phantom with two lesions.

## II. Theory

The particle velocity $\boldsymbol{V}$ within a fully reverberant shear wave field is described as [17, 18] as

$$\boldsymbol{V}(\boldsymbol{\varepsilon}, t) = \sum_{q,l} \widehat{\boldsymbol{n}}_{ql} \, v_{ql} \, e^{i\left(k\widehat{\boldsymbol{n}}_q \cdot \boldsymbol{\varepsilon} - \omega_0 t\right)}. \tag{1}$$

where $\boldsymbol{\varepsilon}$ signifies the position vector, $t$ is time, $\omega_0$ is the angular frequency, and the indices $q$ and $l$ correspond to realizations of the random unit vectors $\widehat{\boldsymbol{n}}_q$ and $\widehat{\boldsymbol{n}}_{ql}$, respectively. The vector $\widehat{\boldsymbol{n}}_q$ denotes the random direction of wave propagation, $\widehat{\boldsymbol{n}}_{ql}$ indicates a random unit vector representing the direction of particle motion and $v_{ql}$ represents an independent, identically distributed random variable signifying the magnitude of the particle velocity within a given realization. In transverse shear wave fields, the direction of wave propagation is orthogonal to the particle motion, indicating that $\widehat{\boldsymbol{n}}_{ql} \cdot \widehat{\boldsymbol{n}}_q = 0$. However, the wave propagation and particle motion are in the same direction for compression waves, which implies $\widehat{\boldsymbol{n}}_{ql} \cdot \widehat{\boldsymbol{n}}_q = 1$.

In the standard autocorrelation technique, the autocorrelation of the $z$-directed velocity field ($B_{V_z V_z}$) is conventionally computed in both space and time as described in [19] as

$$B_{V_z V_z}(\Delta\boldsymbol{\varepsilon}, \, \Delta t) = E\{V_z(\boldsymbol{\varepsilon}, t)V_z^*(\boldsymbol{\varepsilon} + \Delta\boldsymbol{\varepsilon}, t + \Delta t)\} \tag{2}$$

where $\Delta\boldsymbol{\varepsilon}$ and $\Delta t$ represent the small difference in position vector and time, respectively, $E$ signifies an ensemble average, and the asterisk (*) indicates the complex conjugate. In practice, this equation takes the estimated particle velocities from an imaging system as a function of space and time. The presence of any unwanted compression waves in the tissue adds a long wavelength term to the autocorrelation equation. The estimated SWS in this condition is affected by the low spatial frequency compression waves. In order to overcome this issue and minimize the influence of compression waves and whole tissue motion, we propose to compute the autocorrelation of the quantity $V_z(\boldsymbol{\varepsilon} - \Delta\boldsymbol{\varepsilon}) - V_z(\boldsymbol{\varepsilon} + \Delta\boldsymbol{\varepsilon})$ instead of the autocorrelation of only the velocity field $V_z$. The subtraction of particle velocities between neighboring particles effectively cancels out the contribution of compression waves with large wavelengths, leaving only the SWS component. Thus, the difference autocorrelation estimator $B_{DA_{V_z V_z}}$ is defined, for simplicity when $\Delta t$ is zero, as follows

$$
\begin{aligned}
B_{DA_{V_z V_z}}(\Delta\varepsilon) &= E\{[V_z(\boldsymbol{\varepsilon} - \Delta\boldsymbol{\varepsilon}) - V_z(\boldsymbol{\varepsilon} + \Delta\boldsymbol{\varepsilon})] \, [V_z(\boldsymbol{\varepsilon} - \Delta\boldsymbol{\varepsilon}) - V_z(\boldsymbol{\varepsilon} + \Delta\boldsymbol{\varepsilon})]^*\} \\
&= E\left\{ \begin{matrix} [V_z(\boldsymbol{\varepsilon} - \Delta\boldsymbol{\varepsilon}) \, V_z^*(\boldsymbol{\varepsilon} - \Delta\boldsymbol{\varepsilon})] + [V_z(\boldsymbol{\varepsilon} + \Delta\boldsymbol{\varepsilon}) \, V_z^*(\boldsymbol{\varepsilon} + \Delta\boldsymbol{\varepsilon})] \\ -[V_z(\boldsymbol{\varepsilon} - \Delta\boldsymbol{\varepsilon}) \, V_z^*(\boldsymbol{\varepsilon} + \Delta\boldsymbol{\varepsilon})] - [V_z(\boldsymbol{\varepsilon} + \Delta\boldsymbol{\varepsilon}) \, V_z^*(\boldsymbol{\varepsilon} - \Delta\boldsymbol{\varepsilon})] \end{matrix} \right\} \\
&= 2\left( \overline{V_z}^2 - B_{V_z V_z}(2\Delta\varepsilon) \right)
\end{aligned} \tag{3}
$$

where $\overline{V_z}^2$ is the ensemble average velocity-squared and the simplified form is based on the fundamental definition of each of the terms, assuming spatially stationary statistics. Note also that any extra motion approximately constant across the autocorrelation window will be canceled by the subtraction in the first bracketed expressions.

In Equation (3), $B_{V_zV_z}$ is the conventional spatial autocorrelation function. The autocorrelation function in different directions depends on the angle $\theta_s$ between the imaging system sensitivity (assumed to be z-directed) and the direction of $\Delta\boldsymbol{\varepsilon}$, defined by Aleman-Castañeda et al. [20] as follows

$$B_{V_zV_z}(\Delta\boldsymbol{\varepsilon}, \Delta t) = 3\overline{V_z}^2\, e^{i\omega_0\Delta t}\left\{\frac{\sin^2\theta_s}{2}\left[j_0(k\Delta\varepsilon) - \frac{j_1(k\Delta\varepsilon)}{k\Delta\varepsilon}\right] + \cos^2\theta_s\,\frac{j_1(k\Delta\varepsilon)}{k\Delta\varepsilon}\right\}, \qquad (4)$$

where $j_0$ is the spherical Bessel function of the first kind of zero order and $j_1$ is the spherical Bessel function of the first kind of first order. In standard baseline autocorrelation estimation, $\Delta\boldsymbol{\varepsilon}$ is assumed to be aligned with one of the Cartesian axes. The angle $\theta_s$ in equation (4) is $\pi/2$ for $\Delta\varepsilon_x$ and $\Delta\varepsilon_y$, and zero for $\Delta\varepsilon_z$. Consequently, the standard baseline autocorrelation functions are defined as

$$B_{V_zV_z}(\Delta\varepsilon_x, \Delta t)\ = \frac{3}{2}\,\overline{V_z}^2\, e^{i\omega_0\Delta t}\left[j_0(k\Delta\varepsilon_x) - \frac{j_1(k\Delta\varepsilon_x)}{k\Delta\varepsilon_x}\right] \qquad (5.a)$$

$$B_{V_zV_z}(\Delta\varepsilon_y, \Delta t)\ = \frac{3}{2}\overline{V_z}^2\, e^{i\omega_0\Delta t}\left[j_0(k\Delta\varepsilon_y) - \frac{j_1(k\Delta\varepsilon_y)}{k\Delta\varepsilon_y}\right] \qquad (5.b)$$

$$B_{V_zV_z}(\Delta\varepsilon_z, \Delta t)\ =\ 3\overline{V_z}^2\, e^{i\omega_0\Delta t}\,\frac{j_1(k\Delta\varepsilon_z)}{k\Delta\varepsilon_z} \qquad (5.c)$$

To enhance the robustness of the autocorrelation estimator, Asemani et al. [21] introduced the angular integral autocorrelation (AIA) that involves computing the angular integral of the autocorrelation function within a two-dimensional plane spanning from 0 to $2\pi$, resulting in the expression for the $xz$ plane as follows

$$B_{AIA_{xz}}(\Delta\rho, \Delta t) = B_{AI_{yz}}(\Delta\rho, \Delta t) = \frac{3}{4}\overline{V_z}^2\, e^{i\omega_0\Delta t}\left[j_0(k\Delta\rho) + \frac{j_1(k\Delta\rho)}{k\Delta\rho}\right] \qquad (6)$$

where $\Delta\rho$ represents the one-dimensional lag in the autocorrelation argument subsequent to integration around $\theta_s$. For the $xy$ plane we have:

$$B_{AIA_{xy}}(\Delta\rho, \Delta t) = \frac{3}{2}\,\overline{V_z}^2\, e^{i\omega_0\Delta t}\left[j_0(k\Delta\rho) - \frac{j_1(k\Delta\rho)}{k\Delta\rho}\right] \qquad (7)$$

The difference autocorrelation for the standard baseline autocorrelation estimator can be obtained by assuming $\Delta t = 0$ and substituting Equation (5) in Equation (3) to yield

$$B_{DA_z}(\Delta\varepsilon_x) = 2\left(\overline{V_z}^2 - \frac{3}{2}\,\overline{V_z}^2\left[j_0(2k\Delta\varepsilon_x) - \frac{j_1(2k\Delta\varepsilon_x)}{2k\Delta\varepsilon_x}\right]\right) \qquad (8.a)$$

$$B_{DA_z}(\Delta\varepsilon_y)\ =\ 2\left(\overline{V_z}^2 - \frac{3}{2}\,\overline{V_z}^2\left[j_0(2k\Delta\varepsilon_y) - \frac{j_1(2k\Delta\varepsilon_y)}{2k\Delta\varepsilon_y}\right]\right) \qquad (8.b)$$

$$B_{DA_z}(\Delta\varepsilon_z)\ =\ 2\left(\overline{V_z}^2 - 3\overline{V_z}^2 e^{i\omega_0\Delta t}\,\frac{j_1(2k\Delta\varepsilon_z)}{2k\Delta\varepsilon_z}\right) \qquad (8.c)$$

By substituting Equations (6) or (7) in Equation (3), the integrated difference autocorrelation (IDA) for different autocorrelation planes can be obtained as

$$B_{IDA_{xz}}(\Delta\rho) = B_{IDA\ yz}(\Delta\rho) = 2\left(\overline{V_z}^2 - \frac{3}{4}\overline{V_z}^2\left[j_0(2k\Delta\rho) + \frac{j_1(2k\Delta\rho)}{2k\Delta\rho}\right]\right) \qquad (9)$$

$$B_{IDA_{xy}}(\Delta\rho) = 2\left(\overline{V_z}^2 - \frac{3}{2}\,\overline{V_z}^2\left[j_0(2k\Delta\rho) - \frac{j_1(2k\Delta\rho)}{2k\Delta\rho}\right]\right) \tag{10}$$

Figure 1(a) displays the autocorrelation curves for the $z$-directed simple autocorrelation functions and the AIA function in the $xz$ plane. As depicted in this figure, the curves for the autocorrelation functions ($B_{V_zV_z}(\Delta\varepsilon_x)$ and $B_{V_zV_z}(\Delta\varepsilon_z)$) are distinct, and the AIA function is consistently positioned between them. The maximum occurs at zero lag for all autocorrelation functions. Figure 1(b) presents $z$-directed difference autocorrelation curves of $\Delta\varepsilon_x$ and $\Delta\varepsilon_y$, along with the IDA curve in the $xz$ plane. The IDA curve consistently falls between the simple difference autocorrelation curves. At zero lag, the value for the difference autocorrelation functions is zero. In practice, a curve fit of measured data to Equations (9,10) is necessary. Figure 1(b) is utilized to derive a best fit estimate of wavenumber $k$, thereby determining SWS from the physics definition $k = \omega/C_s$ and where $\omega$ is the applied radial frequency and $C_s$ is taken in this context as equal to the SWS.

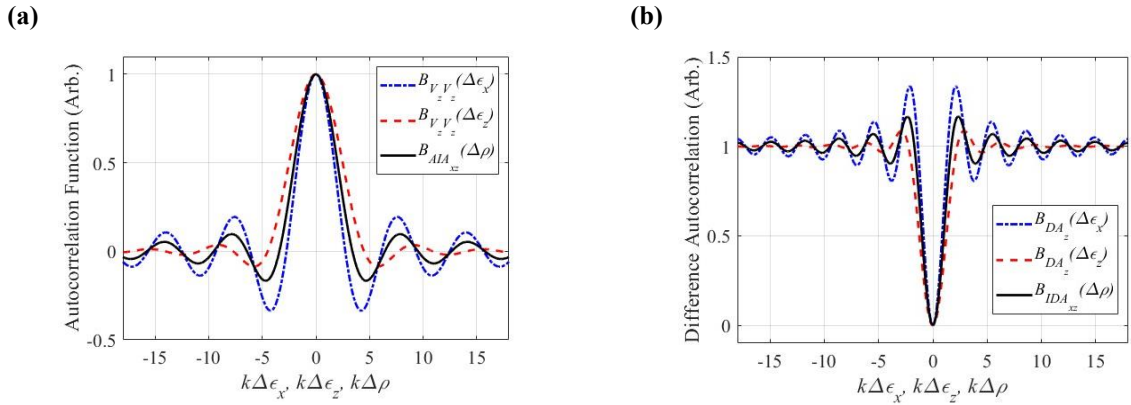

**(a)** **(b)**

**Fig. 1. Autocorrelation and difference autocorrelation curves: (a) curves for the $z$-directed simple autocorrelation functions and the AIA function in the $xz$ plane, (b) curves for $z$-directed difference autocorrelation functions of $\Delta\varepsilon_x$ and $\Delta\varepsilon_y$, along with the IDA function in the $xz$ plane.**

## III. K-WAVE ELASTOGRAPHY SIMULATION

The effectiveness of the proposed IDA approach was evaluated using the k-Wave simulation toolbox in MATLAB (The MathWorks, Inc. Natick, MA, USA, version 2022b) [22] of a stiff branching cylinder in a soft background. Both the background medium and the y-shaped inclusion were modeled as uniform isotropic materials, with SWS of 1 m/s and 2 m/s, respectively. The simulation geometry and the defined SWS properties for the y-shaped cylinder and the background are presented in Figure 2(a). Further details regarding this simulation can be found in [21]. A fully reverberant shear wave field at a frequency of 200 Hz was created by the application of multiple shear wave point sources around the outer boundary. However, for this study the resulting shear wave field was purposely superimposed with compression waves, modeled as a long wavelength phasor of the same temporal frequency and approximately the same amplitude. The velocity field was specifically measured along the z-axis to mimic realistic scanning conditions in elastography. Figure 2(b) presents an unfiltered phase map of the wave field at 200 Hz, revealing the reverberant shear wave field superimposed onto a longer spatial scale of compression waves. The motion movie of the shear wave field is available in the supplementary materials.

A square autocorrelation window 1.5 cm in size was utilized in all simulation measurements to estimate the SWS. A central 3D region of interest was selected to reduce the memory and computational costs. Figure 2(c) displays the SWS map estimated by applying the IDA on the unfiltered contaminated wave field, where the y-shaped inclusion is clearly highlighted. The average SWS in the background was estimated to be 0.98 m/s (i.e., 2% error). The SWS map estimated by applying the AIA on the unfiltered contaminated wave field is illustrated in Figure 2(d). Due to the presence of compression waves in the shear wave field, SWS estimation using conventional methods is challenging, and the y-shaped inclusion is not discernible against the background. The average SWS in the background was estimated to be 3.41 m/s (i.e., 241% error).

Although AIA fails in estimating SWS when applied on contaminated wave fields, effective filtering techniques can enhance its efficiency and accuracy. However, filtering contaminated wave fields is not always straightforward and can be quite challenging, when the range of SWS is not well known *a priori*. One effective filtering method to eliminate the compression waves is the 2D bandpass filter, applied in the Fourier domain by specifying upper and lower bands. The 2D bandpass filter eliminates the waves with a larger spatial wavelength (small wavenumber) as well as smaller wavelengths (largest wavenumbers), however specification of the proper cutoffs requires careful tuning in practice.

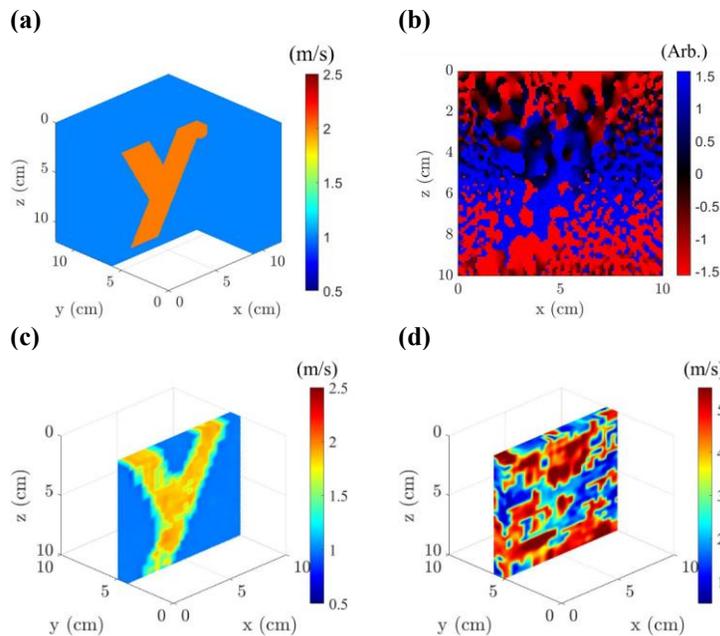

**Fig. 2.** k-Wave elastography simulation: (a) simulation geometry and defined SWS properties for the y-shaped inclusion and the background (b) unfiltered phase map of the contaminated wave field at the frequency of 200 Hz, demonstrating both small scale shear wave patterns and larger scale compression wave pattern, (c) SWS map estimated by applying IDA on unfiltered contaminated wave field, (d) SWS map estimated by applying AIA on the unfiltered contaminated wave field.

In Figure 3(a), the phase map of 2D spatial Fourier transform of the k-Wave displacement field at 200 Hz is depicted. The central red zone signifies the presence of high-energy, low spatial frequency waves or compression waves within the wave field. Figure 3(b) displays the 2D bandpass filter defined for k-Wave elastography, presenting lower (minimum SWS) and upper (maximum SWS) bands. Following numerous trials using different lower and upper bands, the optimal outcome was achieved, with the lower band designated as $C_{s_{min}} = 0.54$ m/s and the upper band as $C_{s_{max}} = 3.89$ m/s. Figure 3(c) depicts the phase map of shear wave field following 2D bandpass filtering at 200 Hz. In this case with perfect *a priori* information, the compression waves are eliminated and a fully reverberant shear wave field is obtained. Employing AIA on the filtered phase map of Figure 3(c) led to the estimation of the SWS map, presented in Figure 3(d). The average SWS in the background was estimated to be 0.98 m/s (i.e., 2% error).

## IV. ULTRASOUND ELASTOGRAPHY OF BREAST PHANTOM

In order to explore the effectiveness of the difference autocorrelation method, a dataset of ultrasound elastography experiments on a CIRS breast phantom (model 509, CIRS Inc., Norfolk, Virginia, USA) was utilized. The CIRS breast phantom mimics the breast tissue characteristics and contains several lesions of different sizes. Four mechanical vibration sources were used in the elastography experiment to generate a reverberant shear wave field. The dataset contains the displacement fields of reverberant shear wave fields at three different excitation frequencies including 900 Hz, 600 Hz, and 400 Hz. Detailed descriptions of the experiments can be found in [16] and [21].

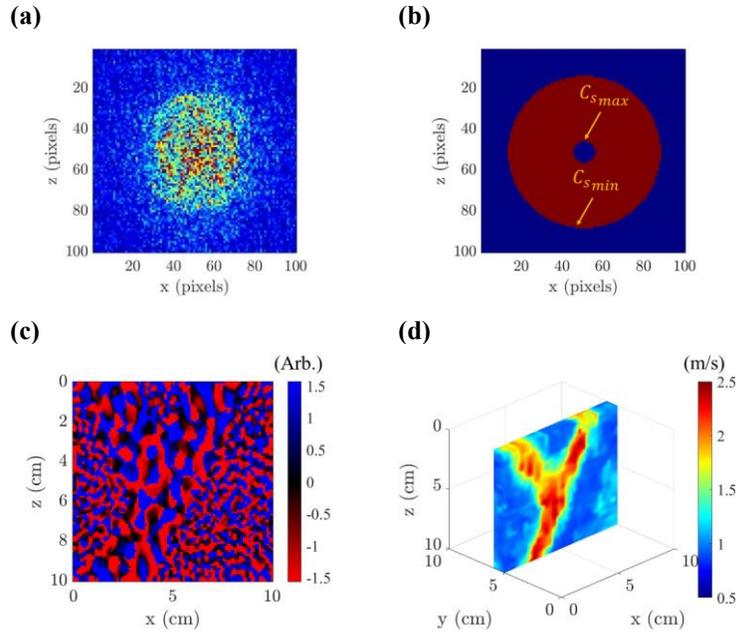

**Fig. 3. Process of 2D bandpass filtering and SWS estimation using AIA in k-Wave simulation: (a) phase map of the 2D spatial Fourier transform of the k-Wave displacement field at the frequency of 200 Hz, (b) 2D bandpass filter in the Fourier domain used to eliminate compression waves, (c) phase map of the shear wave field after 2D bandpass filtering at the frequency of 200 Hz, (d) SWS map estimated by applying AIA on the filtered phase map of the shear wave field at the frequency of 200 Hz.**

The lesion in the SWS map is effectively highlighted against the uniform background in Figure 4 (c), and the average SWS using IDA was estimated to be 2.09 m/s in the background at 900 Hz. Similarly, the IDA approach was applied to the elastography displacement field at excitation frequencies of 600 Hz and 400 Hz, and consistent outcomes were obtained. The average SWS in the background using IDA was determined to be 2.07 m/s at 600 Hz and 2.10 m/s at 400 Hz. Ormachea and Parker [16] reported an average SWS of 2.27 m/s in various frequencies including 900 Hz, 600 Hz, and 400 Hz, while Asemani *et al.* [21] reported a slightly lower value of 2.26 m/s.

Figure 4(a) displays the B-mode ultrasound scan of the breast phantom revealing the presence of a lesion. Figure 4(b) presents the unfiltered phase map of the wave field at a frequency of 900 Hz and the short-wavelength reverberant shear wave field is superimposed with a long-wavelength (low spatial frequency) compression wave. The motion movie of the shear wave field is available in the supplementary materials. Applying the IDA to the unfiltered contaminated wave field of Figure 4(b) yields the SWS map depicted in Figure 4(c). Figure 4 (d) displays a SWS map estimated by applying AIA on the unfiltered contaminated wave field. A 7.5 mm square autocorrelation window was used for all breast phantom ultrasound elastography measurements.

In contrast, while AIA proves to be a robust approach for estimating SWS in the presence of noise and reverberance [21], it fails to accurately estimate SWS in a shear wave field superimposed with compression waves. As depicted in Figure 4(d), AIA yielded significantly higher SWS estimates due to the presence of low spatial frequency and long wavelength compression waves. The average SWS in the background using

AIA was estimated to be 9.80 m/s at 900 Hz which is far from the ground truth value. Furthermore, the lesion is not visible in the SWS map estimated using AIA.

Figure 5(a) shows the phase map of 2D spatial Fourier transform of the breast phantom displacement field at 900 Hz. The red zone at the center indicates that there are high energy, low spatial frequency waves or compression waves in the wave field. Figure 5(b) presents the 2D bandpass filter with specified lower and upper bands.

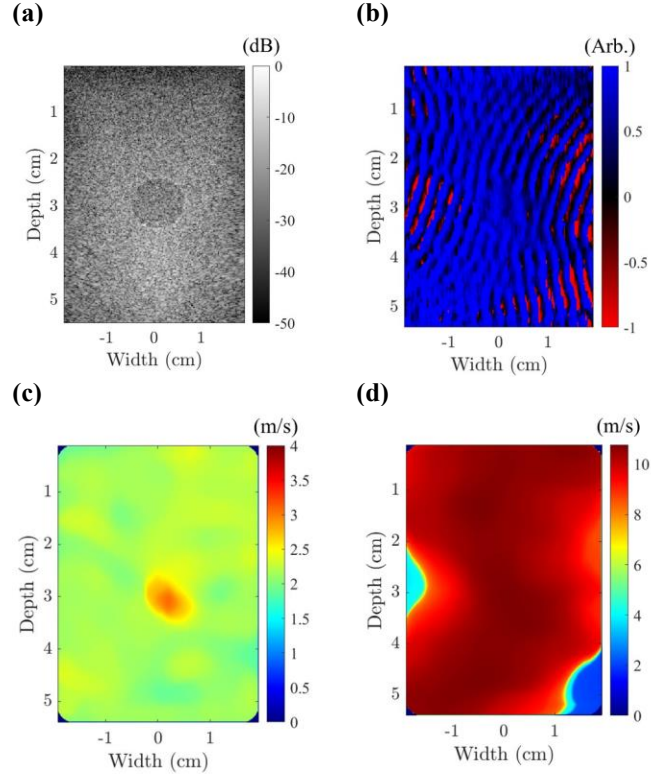

**Fig. 4. Ultrasound elastography of a breast phantom with a lesion: (a) B-mode ultrasound scan of the phantom, (b) unfiltered phase map of the contaminated wave field at the frequency of 900 Hz, (c) SWS map estimated by applying IDA on the unfiltered contaminated wave field, (d) SWS map estimated by applying AIA on the unfiltered contaminated wave field.**

After several attempts with different lower and upper bands, the best results for the frequency of 900 Hz were achieved with the lower band set to $C_{s_{min}} = 0.30$ m/s and the upper band set to $C_{s_{max}} = 6.79$ m/s. The phase map of the displacement field after the 2D bandpass filtering at 900 Hz is displayed in Figure 5(c). Applying AIA to the filtered phase map of Figure 5(c) resulted in the estimated SWS map shown in Figure 5(d). The lesion is highlighted in the SWS map with the average SWS in the background estimated to be 2.45 m/s (i.e., 8% error).

## V. Ultrasound elastography of liver-kidney

In this section, the effectiveness of the developed IDA estimator is explored using a dataset of ultrasound elastography focusing on the human liver-kidney region. The dataset contains B-mode ultrasound scans and displacement field data captured from fully reverberant shear waves across a diverse spectrum of excitation frequencies, including 702 Hz, 585 Hz, 468 Hz, 351 Hz, 234 Hz, and 117 Hz. This ultrasound elastography experiment was performed under the requirements of informed consent of the South-woods Imaging Clinical Institutional Review Board. Further details regarding the *in vivo* experiment can be found in [16] and [21].

The B-mode ultrasound scan of an anatomical view of the liver and kidney is illustrated in Figure 6(a). The scan reveals both tissues, the liver (zone L) depicted by distinct darker zones, and the kidney (zone K) depicted by a rounded region. Figure 6(b) represents the unfiltered phase map of the liver-kidney wave field at a frequency of 702 Hz, showcasing the reverberant shear wave field superimposed with the low spatial frequency compression waves. The motion movie of the shear wave field, available in the supplementary materials, highlights the effect of compression waves on the shear wave field.

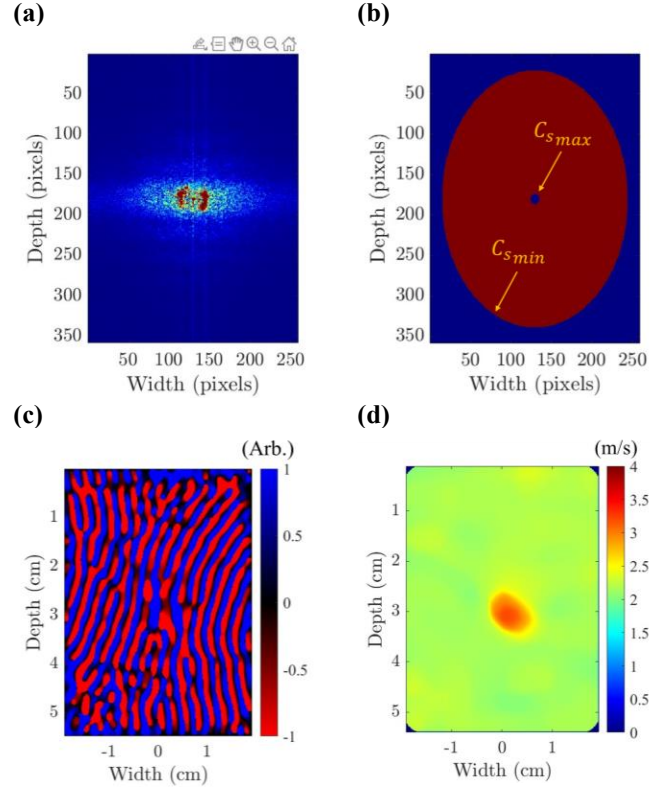

**Fig. 5. Process of 2D bandpass filtering and SWS estimation in breast phantom ultrasound elastography dataset: (a) phase map of 2D spatial Fourier transform of displacement field at the frequency of 900 Hz, (b) 2D bandpass filter in the Fourier domain used to eliminate compression waves, (c) phase map of the shear wave field after 2D bandpass filtering at the frequency of 900 Hz, (d) SWS map estimated by applying AIA on the filtered phase map of the shear wave field at the frequency of 900 Hz.**

**TABLE 1. SWS IN THE BACKGROUND ESTIMATED BY APPLYING IDA AND AIA APPROACHES TO UNFILTERED CONTAMINATED WAVE FIELDS OF THE BREAST PHANTOM ULTRASOUND ELASTOGRAPHY ACROSS DIFFERENT EXCITATION FREQUENCIES**

| Excitation frequency | SWS (m/s) in Breast using IDA | Estimation error using IDA | SWS (m/s) in Breast using AIA | Estimation error using IDA |
|---|---|---|---|---|
| 900 Hz | 2.09 ± 0.11 | 8% | 9.80 ± 1.73 | 332% |
| 600 Hz | 2.07 ± 0.09 | 9% | 1.74 ± 0.09 | 23% |
| 400 Hz | 2.10 ± 0.10 | 7% | 1.65 ± 0.10 | 27% |

All measurements in ultrasound elastography of the liver-kidney region utilized a square autocorrelation window 18.5 mm in size. Figure 6(c) displays the SWS map estimated using IDA on the unfiltered contaminated wave field of liver-kidney. The average SWS was estimated to be 2.57 m/s in the liver and 3.65 m/s in the kidney. The regions for measuring the average SWS are indicated by two dashed rectangles in Figure 6(c). Figure 6(d) showcases the estimated SWS map obtained from the IDA overlaid on the B-mode ultrasound scan of the liver-kidney region. Both the liver region and the kidney region are visualized in the IDA SWS map.

Figure 6(e) presents the SWS map estimated using AIA on the unfiltered contaminated wave field at a frequency of 702 Hz. The liver and kidney regions are not discernible in this SWS map, and the average SWS in the liver and kidney were estimated to be 9.32 m/s and 7.14 m/s, respectively. Asemani *et al.* [21] reported average SWS values of 2.69 m/s in the liver and 3.98 m/s in the kidney. As expected, unlike IDA, the AIA approach fails to accurately estimate SWS in ultrasound elastography of the liver-kidney when the shear wave field is superimposed with compression waves. The IDA estimator was applied to the unfiltered wave field at other excitation frequencies, including 585 Hz, 468 Hz, 351 Hz, 234 Hz, and 117 Hz, yielding consistent results (see Table 2).

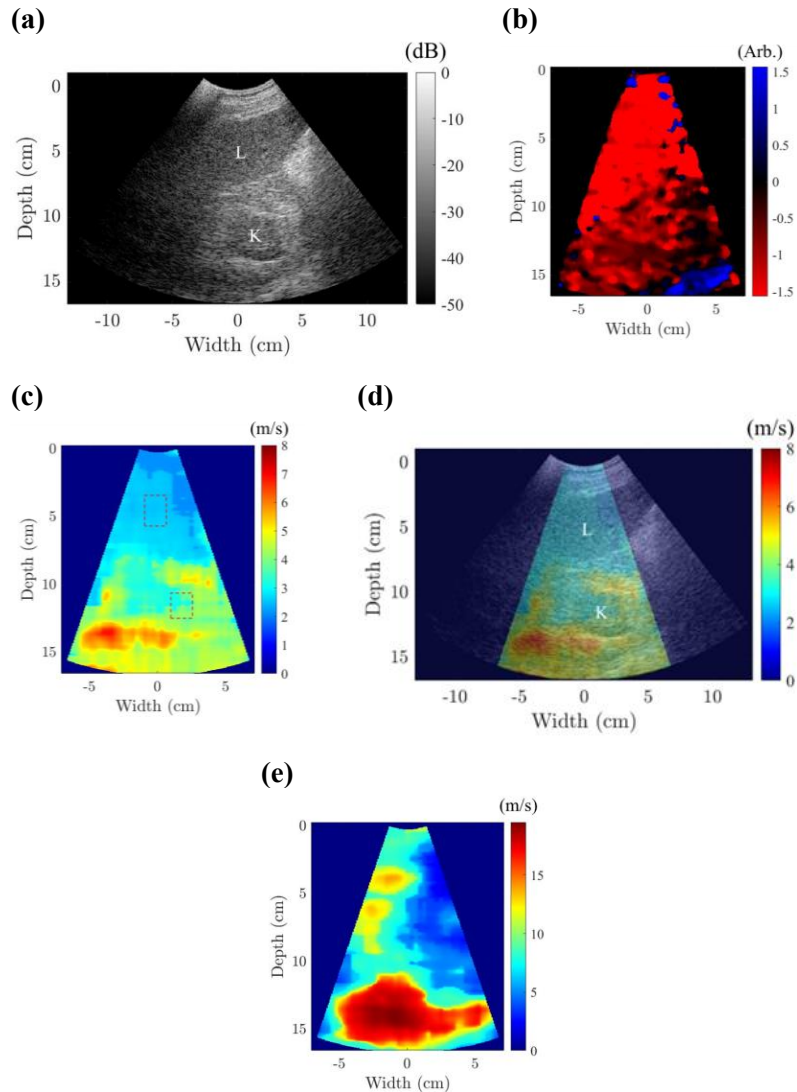

**Fig. 6. Ultrasound elastography of a human liver-kidney region: (a) B-mode ultrasound scan of the liver-kidney, (b) unfiltered phase map of the contaminated wave field at the frequency of 702 Hz, (c) SWS map estimated by applying IDA on the unfiltered contaminated wave field, (d) estimated SWS map obtained from the IDA overlaid on the B-mode ultrasound scan of the liver-kidney region, (e) SWS map estimated by applying AIA on the unfiltered contaminated wave field.**

Figure 7(a), depicts the phase map of 2D spatial Fourier transform of the liver-kidney displacement field at the frequency of 702 Hz. The red region in the center indicates the presence of compression waves in the wave field. The 2D bandpass filter used to eliminate compression waves is presented in Figure 7(b), and the phase map of the displacement field after the 2D bandpass filtering is shown in Figure 7(c). Figure 7(d), illustrates the SWS map estimated using AIA on the filtered displacement field. The liver and kidney regions

are highlighted in the SWS map and the average SWS is estimated to be 2.69 m/s in the liver and 3.98 m/s in the kidney.

**TABLE 2. SWS IN THE LIVER AND THE KIDNEY ESTIMATED BY APPLYING IDA AND AIA APPROACHES TO UNFILTERED CONTAMINATED WAVE FIELDS OF LIVER-KIDNEY ULTRASOUND ELASTOGRAPHY ACROSS DIFFERENT EXCITATION FREQUENCIES**

| Excitation frequency | SWS (m/s) in liver using IDA | SWS (m/s) in liver using AIA | SWS (m/s) in kidney using IDA | SWS (m/s) in kidney using AIA | SWS ratio using IDA | SWS ratio using AIA |
|---|---|---|---|---|---|---|
| 702 Hz | $2.57 \pm 0.04$ | $9.32 \pm 2.33$ | $3.65 \pm 0.39$ | $7.14 \pm 1.98$ | $1.42 \pm 0.15$ | $0.77 \pm 0.29$ |
| 585 Hz | $2.20 \pm 0.13$ | $7.41 \pm 1.84$ | $3.31 \pm 0.12$ | $15.11 \pm 0.44$ | $1.50 \pm 0.10$ | $2.04 \pm 0.51$ |
| 468 Hz | $1.78 \pm 0.34$ | $1.85 \pm 0.18$ | $2.72 \pm 0.07$ | $11.04 \pm 0.57$ | $1.53 \pm 0.29$ | $5.97 \pm 0.66$ |
| 351 Hz | $1.21 \pm 0.10$ | $6.08 \pm 2.42$ | $1.74 \pm 0.25$ | $9.06 \pm 0.16$ | $1.44 \pm 0.24$ | $1.49 \pm 0.59$ |
| 234 Hz | $0.78 \pm 0.08$ | $6.38 \pm 0.20$ | $1.23 \pm 0.10$ | $6.90 \pm 0.02$ | $1.58 \pm 0.21$ | $1.08 \pm 0.03$ |
| 117 Hz | $0.41 \pm 0.03$ | $1.58 \pm 0.18$ | $0.65 \pm 0.11$ | $3.45 \pm 0.03$ | $1.59 \pm 0.29$ | $2.18 \pm 0.25$ |

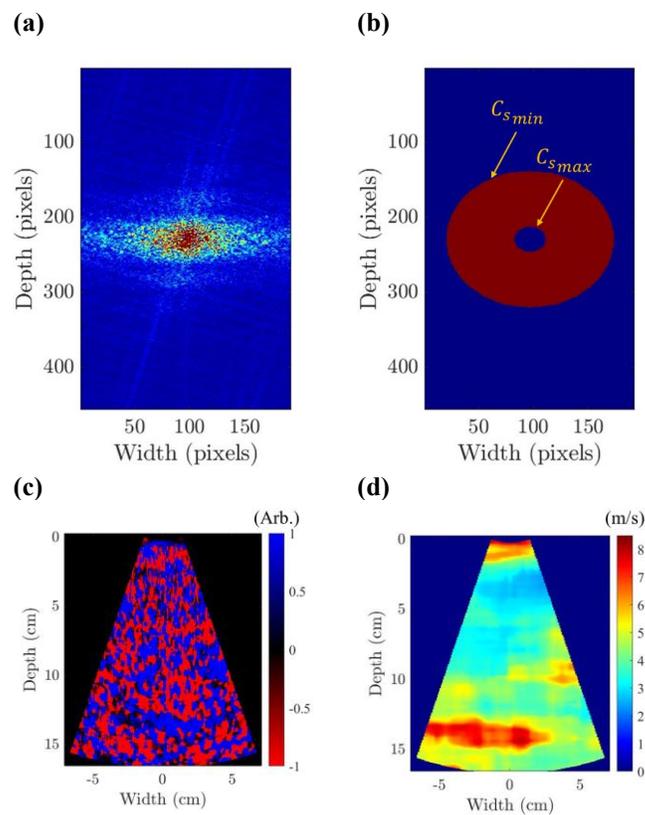

**Fig. 7. Process of 2D bandpass filtering and SWS estimation in liver-kidney ultrasound elastography dataset: (a) phase map of the 2D spatial Fourier transform of the displacement field at the frequency of 702 Hz, (b) 2D bandpass filter in the Fourier domain used to eliminate compression waves, (c) phase map of the shear wave field after 2D bandpass filtering at the frequency of 702 Hz, (d) SWS map estimated by applying AIA on the filtered phase map of the shear wave field at the frequency of 702 Hz.**

## VI. MAGNETIC RESONANCE ELASTOGRAPHY OF BRAIN PHANTOM

In order to assess the effectiveness of the developed IDA approach across different modalities and excitation scenarios, the performance of this approach was evaluated using an MRE dataset of a brain phantom containing two lesions. Figure 8(a) presents a B-mode MRI scan of a cross-section in the *xy* plane of the brain phantom. The brain phantom included two spherical gelatin lesions, with diameters of 18 mm and 12 mm, respectively. A pneumatic mechanical actuator was used to generate the reverberant shear wave field within the phantom. In Kabir *et al.* [23], the background shear modulus for the brain phantom is reported as $3.34 \pm 0.04$ kPa, which corresponds to SWS of 1.83 m/s. The details of the experiments can be found in [23].

The unfiltered phase map of the Y-motion wave field at a frequency of 200 Hz is illustrated in Figure 8(b). The motion movie for wave field is provided in the supplementary materials. The Y-motion demonstrates superimposition with large wavelength and low spatial frequency compression waves, rendering the shear wave pattern indiscernible. Figure 8(c) depicts the SWS map estimated by applying the IDA estimator on the unfiltered contaminated wave field of Y-motion at the frequency of 200 Hz. A 10.7 mm square autocorrelation window was applied in all MRE measurements. Lesions are successfully highlighted in this SWS map, and the average background SWS is estimated to be 2.18 m/s (i.e., 19% error). The MRE result demonstrates the IDA's effectiveness in estimating SWS in the highly contaminated wave field, without requiring any filtering. This validates the robustness of the IDA estimator in the presence of compression waves.

Figure 8(d) shows the SWS map estimated by applying AIA on the Y-motion displacement field. The average SWS in the background using AIA on unfiltered wave field is estimated to be 4.89 m/s. Lesions are not visualized in the SWS map and the AIA approach fails to estimate SWS in the unfiltered contaminated wave field of Y-motion, as expected.

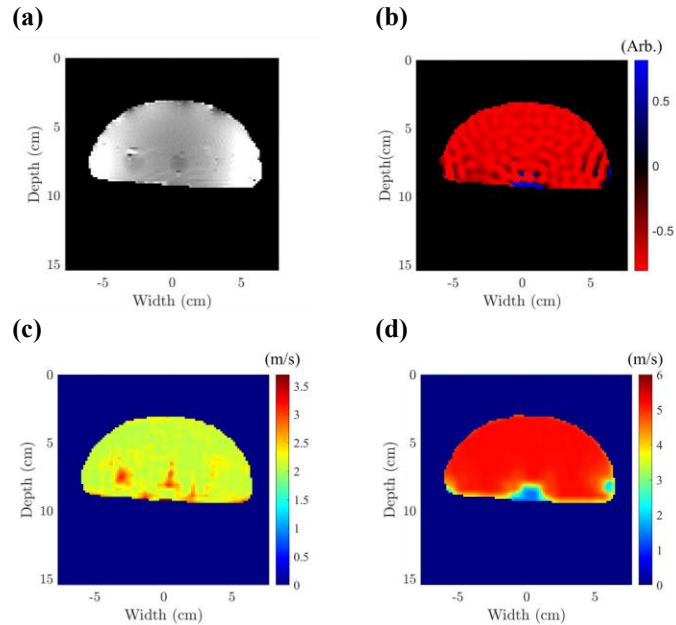

**Fig. 8. MRE of a brain phantom with two lesions at the frequency of 200 Hz: (a) B-mode MRI scan of the phantom, (b) unfiltered phase map of contaminated wave field of Y-motion, (c) SWS map estimated by applying IDA on the unfiltered contaminated wave field of Y-motion, (d) SWS map estimated by applying AIA on the unfiltered contaminated wave field of Y-motion.**

As discussed in Section 3, the effectiveness of the AIA approach in the presence of compression waves may be enhanced when used alongside appropriate filtering techniques. Figure 9 illustrates the filtering process in the MRE dataset. In Figure 9(a), the phase map of the 2D spatial Fourier domain of the MRE displacement field at a frequency of 200 Hz is presented. The prominent red zone in the center indicates the presence of strong low spatial frequency waves or compression waves contaminating the shear wave field. This extensive red zone signifies the strength and wide spatial frequency range of the compression waves.

Figure 9(b) showcases the 2D bandpass filter tailored for the MRE dataset in the Fourier domain. After experimenting with different combinations of lower and upper bands, optimal results were achieved with the lower band set to $C_{s\,min} = 0.68$ m/s and the upper band set to $C_{s\,max} = 2.80$ m/s. Figure 9(c) displays the phase map of the shear wave field after 2D bandpass filtering at the frequency of 200 Hz. The impact of the compression wave is notably reduced, and the reverberant shear wave pattern becomes discernible. The motion movie of the shear wave field after filtering is provided in the supplementary materials.

Figure 9(d) exhibits the estimated SWS map using AIA on a filtered shear wave field of Y-motion at the frequency of 200 Hz. The average SWS in the background estimated using AIA is 2.30 m/s (i.e., 26% error), and the lesions are not visualized in this SWS map. This illustrates that even after filtering and mitigating the effects of compression waves, AIA may fail in visualizing lesions and estimating SWS, particularly when the shear wave field is superimposed with compression waves exhibiting a wide spatial frequency range. In contrast, IDA consistently proves effective in SWS estimation, even in the presence of strong compression waves across a wide spatial frequency range.

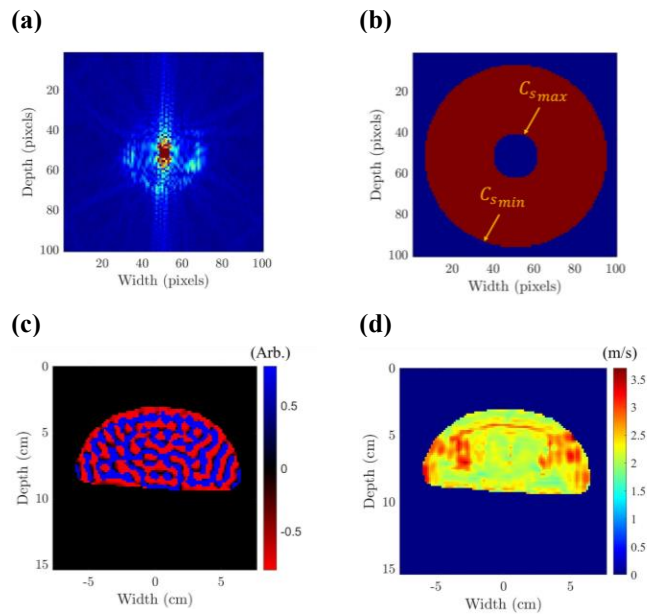

Fig. 9. Process of 2D bandpass filtering and SWS estimation using AIA in MRE data set: (a) phase map of 2D spatial Fourier domain of Y-motion displacement field at the frequency of 200 Hz, (b) 2D bandpass filter in the Fourier domain used to eliminate compression waves, (c) phase map of the Y-motion shear wave field after 2D bandpass filtering at the frequency of 200 Hz, (d) SWS map estimated by applying AIA on the filtered shear wave field of Y-motion at the frequency of 200 Hz.

## VII. DISCUSSION AND CONCLUSION

This study describes the novel difference autocorrelation estimator for SWS estimation in the presence of compression waves. By computing the angular integral of a spatial-difference autocorrelation, the IDA estimator is derived. Through comprehensive evaluations using k-Wave simulations and experimental data from ultrasound elastography and MRE, we have demonstrated the effectiveness and robustness of IDA in accurately estimating SWS across various tissue types and imaging modalities where the presence of long wavelength compression waves complicates the calculation of shear wave properties. There are several advantages to the newer IDA approach. There is no need to have 3D vector data to implement the vector curl operator. Further, careful *a priori* estimates of wavelengths to fine tune a bandpass filter are not required. The IDA is estimated directly from the velocity data by means of the autocorrelation operation in 2D, integrated to a one dimensional (radial average) function of lag.

To assess the effectiveness of the proposed IDA approach, a fully reverberant shear wave field superimposed with low spatial frequency compression waves was introduced in a k-Wave model of a stiff y-shaped cylinder within a uniform soft background. The IDA estimator displayed advanced capabilities in highlighting the y-shaped inclusion and estimating the SWS in background accurately with a 2% error.

In ultrasound elastography of a breast phantom with a lesion, IDA effectively highlighted the lesion and consistently estimated SWS values across different excitation frequencies with less than 9% error, demonstrating its reliability in complex wave fields with imperfect reverberation and compression waves. Similarly, in ultrasound elastography of the liver-kidney region, IDA successfully visualized liver and kidney tissues, providing consistent SWS estimates, even in the presence of compression wave contamination, thus establishing its effectiveness in clinical scenarios.

In the MRE of a brain phantom, IDA showcased its capability to reasonably estimate SWS with 19% error even in highly contaminated wave fields, thereby verifying its reliability across diverse modalities and excitation scenarios. The comprehensive evaluation across multiple modalities in this study underscores the robustness and potential clinical applicability of the IDA approach in shear wave elastography. By effectively mitigating the impact of compression waves, IDA offers a promising solution for precise SWS estimation across several modalities, enhancing the diagnostic capabilities of elastography techniques.

## ACKNOWLEDGMENT

This work was supported by National Institutes of Health grant R21AG070331 and the University of Rochester Center of Excellence in Data Science for Empire State Development, grant number 2089A015. We are grateful to Elastance Imaging, Inc. and Dr. Richard Barr for sharing a scan from their previous studies.